# Potential Role of Agentic Artificial Intelligence in Toxicologic Pathology

**Nasir Rajpoot** (Histofy, UK), **Richard Haworth** (RosettaPath, UK), **Xavier Palazzi** (AstraZeneca, USA), **Alok Sharma** (Labcorp, Madison WI, USA), **Manu Sebastian** (MD Anderson Cancer Center, USA), **Stephen Cahalan** (GSK, UK), **Dinesh S. Bangari** (Sanofi, USA), **Radhakrishna Sura** (Gilead, USA), **James Hartke** (Gilead, USA), **Marco Tecilla** (Roche Pharma Research and Early Development, Switzerland), **Krishna Yekkala** (Johnson & Johnson, USA) **Simon Graham** (Histofy, UK), **Dang Vu** (Histofy, UK), **David Snead** (Histofy, UK), **Mostafa Jahanifar** (Histofy, UK), **Adnan Khan** (Histofy, UK), **Erio Barale-Thomas** (Johnson & Johnson, Belgium)

# Executive Summary

Toxicologic pathology is pivotal to assessing drug safety during preclinical development. However, the process of study reporting remains largely, manual, slow, and fragmented, despite growing demands for speed, consistency, and improved interpretation of human relevance findings.

In June 2025, Histofy hosted a dedicated roundtable discussion during the 44[th] Society of Toxicologic Pathology (STP) Annual Meeting in Chicago. This closed session brought together leading pathologists, CRO scientists, pharmaceutical stakeholders, and AI innovators to examine the capabilities of agentic AI systems for coordinating multiple intelligent agents to accelerate and streamline toxicologic pathology reporting.

Key takeaways included:

- **Need for Efficiency**: There is a clear need to reduce time associated with report generation, data integration, quality assurance and review;

- **Potential of Agentic AI**: Agentic AI systems were seen as potentially useful assistants that could pre-populate study narratives, query internal and external data sources, and augment decision-making without replacing expert judgement. Additionally, they could aid with data analysis and presentation, improve data visualization and comparison of various datasets (anatomic pathology, clinical pathology, body and organ weights, clinical observations, PK/TK);

- **Foundational Pillars**: Trust, traceability, and regulatory alignment emerged as essential pillars for adoption;

- **Practical Path Forward**: Early pilots in non-GLP studies (e.g., discovery-phase, lead optimization, dose-range finding etc) offer a low-risk path to demonstrate practical value.

This white paper synthesizes the discussion and provides a framework for understanding, validating, and adopting agentic AI systems in toxicologic pathology. It is both a roadmap and a call for collaboration across the scientific (pathologists, data scientists, biologists), technical (histo-technicians), and regulatory communities to ensure responsible and scientifically robust use of these tools.

## Abstract

As the volume and complexity of nonclinical toxicology studies continue to increase, toxicologic pathology reporting faces persistent challenges, including fragmented sources of data (e.g., histopathology images, clinical pathology and other study data, adverse effects database, mechanistic literature), variable reporting timelines and heightened regulatory expectations. This white paper examines the emerging role of agentic artificial intelligence (AI) in addressing these issues through coordinated workflow orchestration, data integration, and pathologist-in-the-loop report generation.

Based on a closed-door roundtable held during the 2025 Society of Toxicologic Pathology (STP) Annual Meeting and follow-on discussions, this paper synthesizes the perspectives of leading toxicologic pathologists, toxicologists, and AI developers. It outlines the key pain points in current reporting workflows, identifies realistic near-term use cases for agentic AI, and describes major adoption barriers including requirements for transparency, validation, and organizational readiness. A phased adoption roadmap and pilot design considerations are proposed to help support responsible evaluation and deployment of agentic AI system in nonclinical settings. The paper concludes by emphasizing the need for coordinated efforts across pharmaceutical organizations, CROs, academia, and regulators to establish shared standards, benchmarks, and governance frameworks that will lead to safe, transparent, and trustworthy integration of AI into toxicologic science.

## 1. Introduction

Across the nonclinical safety ecosystem, toxicologic pathology as a discipline is facing tremendous pressure due to increasing study volumes and increasingly larger and more heterogenous datasets under highly compressed timelines **[1-4]**, often while participating in peer review and other related meetings. Pathologists are required to synthesize information from histopathology images, SEND (Standard for Exchange of Nonclinical Data) datasets, clinical pathology and other study data and mechanistic literature. Bringing these diverse information streams into coherent, evidence-based conclusions is both demanding and time-consuming, potentially increasing the risk of oversight and contributing to variability and delays in study reporting. Despite advancements in digital pathology and Artificial Intelligence (AI) for image analysis, the study reporting process remains largely manual **[1]**, with bottlenecks such as slide scanning throughput, data-handling burdens, and assimilation of findings and relevant evidence for report writing. Information often lives in silos: clinical pathology data in one system, histopathology images in another, legacy studies on internal archives, and literature or adverse-event data spread across public databases.

At the same time, the field of AI is undergoing a notable shift. While traditional AI agents are typically designed to automate discrete, individual tasks (such as lesion detection, image classification or data extraction), a new generation of systems known as "agentic AI" is emerging to address end-to-end workflows, either serially or in parallel **[5-9]** – see **Appendix 1 (Glossary)** for a brief description of related terms. These systems integrate multiple specialized agents that work collaboratively: one might gather relevant data, another interpret context, a third synthesize findings across sources, and another draft narrative content, all with real-time orchestration and human-in-the-loop guidance. The result is an intelligent workflow engine that does not simply automate tasks, but coordinates the process from raw inputs to pathologist-reviewed output. A recent study by Ferber *et al*. **[10]** demonstrated the potential of agentic AI systems in oncology, integrating large language models with multimodal tools to autonomously perform data retrieval, image analysis, and clinical reasoning in real-world decision-making scenarios. Similar multi-agent architectures built on large language models can be employed for design of new nanobodies **[11]** or for drug discovery by autonomously integrating biomedical data and accelerating tasks such as literature synthesis, toxicity prediction, and experimental design **[12-14]**. Agentic AI can also be applied to toxicologic pathology, where workflows demand the orchestration of heterogeneous data sources, expert interpretation, and regulatory alignment.

This white paper draws on insights from a roundtable discussion on "Role of Agentic AI for Toxicologic Pathology" held during the Annual Meeting of the Society of Toxicologic Pathology (STP) in Chicago in June 2025. The roundtable participants explored pain points, use cases, adoption barriers, and the future role of intelligent assistants in study reporting. We outline key technical, operational, and regulatory challenges that currently limit integration of agentic AI into toxicologic workflows, several of these – such as data standardization, validation, and infrastructure constraints – are directly highlighted in a recent review of digital pathology adoption in nonclinical settings **[1]**. Additional challenges, such as integration into existing GLP workflows, and broader concerns around transparency, accountability, and regulatory acceptance, also shape the landscape. Addressing these challenges is critical to enable safe, compliant, and scalable deployment of AI agents in preclinical study environments.

To support responsible evaluation and deployment of agentic AI in nonclinical toxicologic pathology, we propose a phased adoption roadmap alongside practical pilot design guidelines. The roadmap outlines incremental stages - from exploratory testing and low-risk applications, to validated integration within regulatory workflows. The pilot framework emphasizes human oversight, traceability, and clear performance metrics, ensuring that early deployments lead to meaningful insights without compromising scientific or regulatory standards.

We surmise that agentic AI could provide meaningful support across several aspects of toxicologic pathology, including: (1) improving access to relevant internal and external information, (2) maintaining the interpretative authority of the pathologist and the overall responsibility over data interpretation and the final report, (3) supporting reproducibility, transparency, and regulatory confidence, and (4) streamlining and accelerating study-report writing. The goal is not to replace expert decision-makers, but to give them better tools for navigating increasingly complex multimodal datasets, allowing them to focus on delivering consistent and evidence-backed scientific judgments.

## 2. Current Pain Points in Toxicologic Pathology Reporting

Despite advances in digital infrastructure and AI for narrow tasks, the core workflow of toxicologic pathology reporting remains manual and time-consuming. The roundtable participants highlighted several persistent challenges that agentic AI systems are well-positioned to address.

### 2.1 Fragmented and Siloed Data

In today's toxicologic pathology workflows, critical study data relevant to the pathologist is often scattered across multiple siloed platforms **[1]**. These include LIMS systems housing SEND datasets and clinical pathology tables, digital slide viewers and image archives, internal legacy reports and spreadsheets, as well as external sources such as scientific literature, FDA databases, and repositories like PubChem. Visualization of data from the LIMS or pathology IMS systems is often limited due to the format of exported files (usually limited to PDF or CSV or Excel formats). Furthermore, integrating these data streams to contextualize findings is often inefficient and relies on manual collation, or informal, interpersonal queries to study directors or research associates – introducing variability, delays, and the risk of overlooking key information. Agentic AI systems can mitigate this by retrieving, linking, and presenting multimodal data in a coherent, context-aware manner.

### 2.2 Time Pressure and Study Bottlenecks

Pathologists are often expected to deliver comprehensive study reports under highly compressed timelines, even in studies involving multiple organs, endpoints, and time points. This pressure can be particularly acute in late-phase or sponsor-facing studies, where deadlines are externally driven. However, histopathology slide review is an inherently time-intensive process, especially in long-term studies such as 2-year carcinogenicity or those involving complex pathology like neurotoxicity.

Since histopathology often sits at the end of study timelines and final report deadlines are rarely moved, delays in earlier phases are often absorbed by compressing the pathology window despite the critical importance of this phase. This may lead to a mismatch between reporting expectations and the realities of slide review and data interpretation. To make matters further challenging, much of the available time is consumed by data handling, cross-referencing, and administrative tasks, leaving limited scope for in-depth interpretation and consistency checks. These workflow constraints, while not universal, may contribute to reporting variability and delays.

## 2.3 Repetition and Lack of Reuse

A significant portion of the reasoning applied in toxicologic pathology reports, such as assessing whether a lesion is background or test article-related, is often repeated, even when highly similar findings have previously been encountered. Most organizations lack systems that allow pathologists to search prior studies by lesion type, study context, or interpretation, making it difficult to leverage institutional memory. In addition, there is often limited ability to summarize patterns across past studies or historical control data and no streamlined way to access those for decision support. This lack of systematic retrieval of institutional knowledge contributes to potential variability, inefficiency, and missed opportunities for consistency.

## 2.4 Risks in Report Quality and Oversight

The roundtable participants expressed several concerns around the practical challenges of toxicologic pathology reporting. Under intense time pressure, there is a risk that subtle but important findings may be overlooked, potentially contributing to false negatives or inconsistencies in interpretation. The quality assurance (QA) feedback process, while essential, can introduce delays to report finalization, especially when iterative reviews are required. In addition, inconsistencies in terminology and reasoning between pathologists can reduce clarity and reproducibility. Complex findings might also be condensed or simplified when access to supporting data is limited, or when reporting timelines restrict deeper investigation. These issues collectively underscore the need for intelligent systems that can support accuracy, consistency, and efficiency in reporting.

## 2.5 Cognitive Load and Human Bottlenecks

The roundtable participants agreed that in current workflows, pathologists often depend on research associates or themselves manually navigate across multiple disconnected systems to gather the information needed to complete their reports. The limited availability of suitable tools for data summarization, data visualization, trend detection, and automated reference retrieval means that highly trained professionals are frequently burdened with low-level synthesis tasks, sometimes leaving insufficient time for multi-source study data interpretation and decision-making.

# 3. Understanding Agentic AI: A New Paradigm for Scientific Workflows

The term *agentic AI* refers to a class of intelligent systems designed to orchestrate multiple AI agents – see **Appendix 1 (Glossary)** for a brief description of technical terms – each with a specific function, towards a shared goal **[5,6]**, for instance the optimization of study reporting workflow or a part thereof. Unlike traditional AI tools that focus on individual tasks (e.g., image classification or data analytics), agentic AI systems consist of multiple agents working in tandem for achieving the shared goal.

## 3.1 Traditional AI vs. Agentic AI

Traditional AI systems are typically designed to automate single, narrowly defined tasks, such as answering a question or counting mitoses on a histology slide. Despite recent developments in AI for pathology image analysis **[15-24]**, these modules (or agents) usually operate as standalone models or tools, with limited contextual awareness. By contrast, agentic AI systems are built to automate entire workflows, not just isolated tasks. They consist of networks of interacting agents, each capable of gathering data, interpreting results, querying references, or drafting reports, all while engaging with users in a dynamic, conversational, and human-in-the-loop manner. While traditional AI tools offer value through speed and task automation, agentic AI systems promise end-to-end workflow efficiency, integrated decision support, and a more coherent synthesis of information across silos. A comparative summary of traditional AI and agentic AI is provided in **Table 1**.

**Table 1**: A Comparison of a Traditional AI tools vs an Agentic AI system

| Feature | Traditional AI Tools | Agentic AI System |
|---|---|---|
| **Scope** | Automates a single task | Automates an entire workflow or a specific workflow segment |
| **Structure** | Standalone model or tool | Network of interacting agents |
| **User interaction** | One-off or repeated input-output batches/cycles | Conversational, dynamic, human-in-the-loop |
| **Examples** | AI counting mitoses or detecting lesions, ChatGPT or CoPilot answering question(s) | AI system interpreting data, querying references, drafting reports |
| **Value Proposition** | Speed, task automation | Workflow efficiency, decision support, integration |
| **Context Awareness** | Operates in isolation with limited context | Continuously gathers and updates context across data sources |
| **Adaptability** | Fixed rules or model output | Dynamically adapts based on user input, data updates, and goals |
| **Outcome** | Answers a question or completes a task | Recommends decision-ready insights or full report draft |

In toxicologic pathology, traditional AI tools may assist with narrow tasks such as flagging abnormal regions on a slide or extracting tables from a SEND file. However, it is the next generation of agentic AI systems that can deliver true end-to-end support with potentially significant gains in reporting time, accuracy, consistency and scientific value. These systems are capable of ingesting protocols and study design metadata, querying relevant internal repositories, and synthesizing trends across data types such as organ weights, anatomic pathology, clinical pathology, body and organ weights, pharmacokinetics (PK) and toxicokinetics (TK) data, and clinical observations.

In addition, agentic AI systems can retrieve supporting references from the literature or FDA adverse event databases and even pre-populate a draft study report for expert review. This holistic, multi-layered capability marks a fundamental shift from task automation to integrated, decision-ready assistance.

## 3.2 How Agentic AI Works in Practice

An agentic AI system in toxicologic pathology could comprise a coordinated set of specialized agents, each contributing to a different stage of the reporting workflow, as shown in **Figure 1**. A ***data ingestion agent*** would read data and images, harmonize inputs from SEND files, LIMS platforms, and associated study metadata to create a unified data foundation. The ***data analytics agent*** could analyze the data for revealing statistically significant findings that may be dose-related and perhaps also analyze WSIs to identify and grade lesions. An ***evidence integration agent*** could retrieve relevant findings, annotations, and precedent cases from scientific publications and regulatory databases. It may also support linkage of observed morphologic changes to known pharmacologic mechanistic pathways, where appropriate. A ***report generation agent*** could then assemble this information into a structured narrative, complete with citations. Finally, a ***QA agent*** would audit the report for internal consistency, completeness, and alignment across datasets. Together, these components enable a more robust, explainable, and efficient reporting process than what traditional tools can offer.

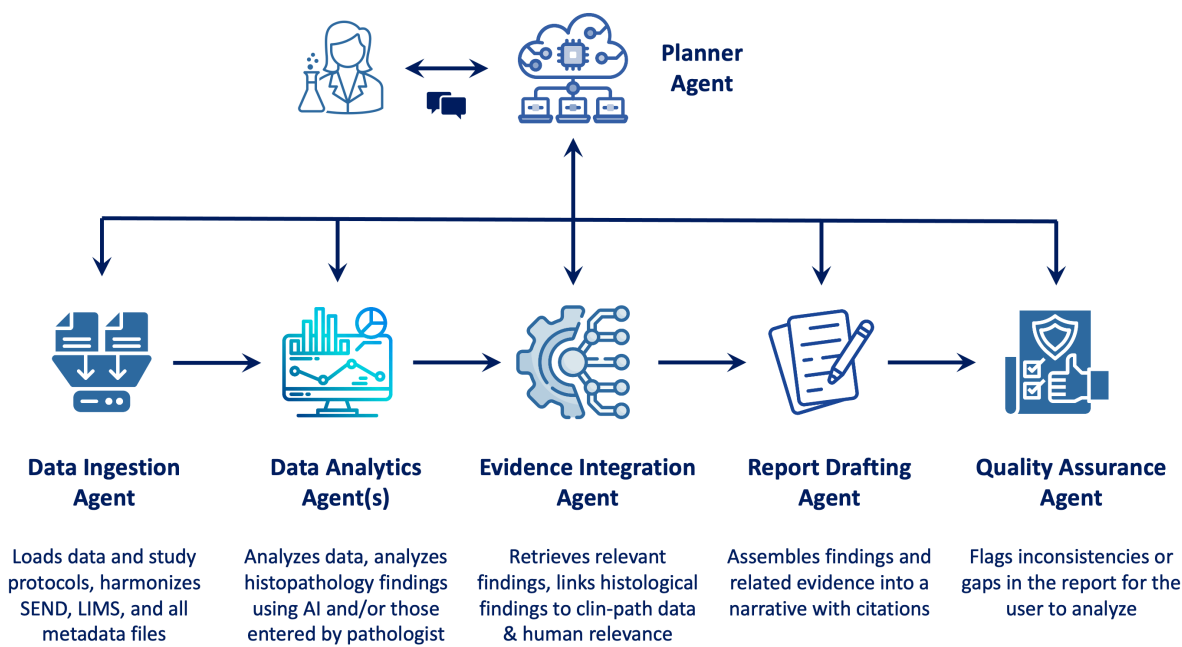

*Figure 1: A sample agentic workflow for toxicologic pathology*

These agents work in coordination, guided by a **_planner agent_**, and engage with the pathologist through a natural language interface. The pathologist remains in control – reviewing, correcting, and approving the output.

### 3.3 The Promise of Agentic AI in Toxicologic Pathology

Agentic AI systems offer the potential to transform toxicologic pathology workflows, streamlining data access, supporting consistent reasoning, and ultimately freeing up expert time to focus on the scientific nuances that matter most in study reporting. These systems can support faster iteration during hypothesis testing and report drafting, enabling more agile and responsive study assessments. By shifting the burden of low-level synthesis and data search to intelligent agents, agentic AI can allow toxicologic pathologists to focus on the scientific and interpretive elements of their role, while maintaining full oversight and editorial control.

## 4. Opportunities for Agentic AI in Toxicologic Pathology

As toxicology studies grow in scale and complexity in terms of multiple data modalities and the range of study endpoints **[1-2]**, the need for intelligent tools that assist - and not replace - expert interpretation is clearer than ever. Agentic AI offers a powerful opportunity to address the challenges outlined in **Section 2** by acting as an intelligent, interoperable assistant across the reporting workflow.

### 4.1 Augmenting, Not Replacing, the Pathologist

Participants consistently emphasized that the adoption of AI in toxicologic pathology should preserve the pathologist's central role in interpreting findings and owning the final report. The true value of agentic AI lies not in replacing expert judgment, but in augmenting it by accelerating routine tasks such as data extraction, citation generation, and summarization. These systems can pre-populate draft content for expert review and correction, allowing pathologists to focus on high-value interpretation. Furthermore, by surfacing relevant references and providing confidence scores, agentic AI can support and enhance interpretive judgments, ultimately strengthening both the quality and defensibility of study reports.

## 4.2 Integrating Internal and External Data Sources

Agentic AI systems are capable of drawing simultaneously from a wide range of data sources. These include internal resources such as study protocols, SEND datasets, and pathology image archives, as well as legacy reports, control databases, and proprietary institutional knowledge bases. In addition, they can access public repositories like PubMed, FDA adverse-event and other approved-drug databases, and PubChem. By synthesizing information across these silos, agentic AI acts as a cross-domain connector, effectively enabling the generation of contextualized, evidence-backed narratives that reflect the full complexity of a study. This integrated approach has the potential to enhance both the efficiency and scientific rigor of toxicologic pathology reporting.

## 4.3 Drafting Study Reports with Citations and Transparency

Participants broadly agreed that one of the most promising near-term applications of agentic AI in toxicologic pathology is the automated generation of first-draft study reports. These drafts would be designed to align with institutional templates and preferred terminology, ensuring consistency with existing reporting standards. An easy win for study report drafts is to accurately pre-populate the Materials and Methods section, where there is room for human error (e.g., specimen list and number of animals examined).

A key feature would be full traceability, clearly indicating which datasets, literature sources, and decision rules informed each section of the draft. Additionally, confidence scores and automated "sanity check" alerts could help enable consistency and flag outliers (e.g., rare or unusual findings) or ambiguities for expert review. Importantly, these drafts are not viewed as final products, but rather as starting points that reduce both the cognitive and clerical load on pathologists, freeing them to focus on interpretation and decision-making while ensuring there is minimal avoidance of critical analysis due to the so-called '*algorithm bias*' **[25]**.

## 4.4 Accelerating QC, Streamlining QA and Consistency Checks

Agentic AI systems also have the potential to play a valuable role in quality control and interpretive support throughout the reporting process. They could assist with the QC review of data tables and incidence rates, missing tissue, flagging discrepancies or missing values early. By applying structured logic, these systems assist with distinguish background findings from treatment-related lesions (e.g., by comparing incidences between groups and correlating histopathological and clinical pathology findings) and assess the adversity and potential human relevance of findings, based on study design and known mechanisms. Such capabilities would enable earlier detection of any reporting inconsistencies, reduce the number of QA-driven revisions, and promote greater consistency in language and reasoning across reports by different reviewers.

## 4.5 Empowering Earlier Phases of R&D

Several participants highlighted that early-phase contexts offer particularly fertile ground for demonstrating the value of agentic AI in toxicologic pathology. These include discovery and lead optimization studies as well as non-GLP exploratory research, where rapid iteration and flexible analysis are often critical. Diverse response domains such as neurotoxicity or reproductive toxicology, which involve complex, nuanced findings, were also seen as strong candidates for early deployment. In these settings, agentic AI can showcase its strengths in usability, time savings, and decision support, without the constraints of formal regulatory reporting. Success in these domains could help build confidence, refine capabilities, and lay the foundation for broader industry adoption.

# 5. Barriers to Adoption

While the promise of agentic AI in toxicologic pathology is clear, roundtable participants identified several practical, cultural, and regulatory hurdles that may need to be addressed for real-world implementation. These barriers do not suggest rejection of the technology, rather they emphasize the need for careful design, transparent operation, and stakeholder collaboration.

## 5.1 Trust, Transparency, and Interpretability

One of the most consistent concerns voiced by participants was trust in the outputs of agentic AI systems. Several cited past experiences with AI tools generating references that seem plausible but are fabricated (similar to '*hallucinations*'), which significantly undermined user confidence. A lack of clear traceability (i.e., visibility into how conclusions were derived and what sources were used) makes users hesitant to rely on AI-generated interpretations, especially in regulated settings. Across the roundtable, there was a strong preference for interpretable AI systems that can explain their reasoning, cite their sources transparently, and allow experts to verify each step in the decision-making process.

In addition, data integrity is governed by the ALCOA++ principles **[26]** originally developed by the FDA and widely adopted across pharma and CROs. These principles require data to be Attributable, Legible, Contemporaneous, Original, and Accurate, with extended expectations for data to also be Complete, Consistent, Enduring, Available, and Traceable. Agentic AI systems that assist with toxicologic pathology reporting may be required to adhere to these expectations if they are to be deployed in GLP-compliant settings. By aligning AI transparency and interpretability with the ALCOA++ framework, developers can better support validation, user trust, and regulatory acceptance.

## 5.2 Validation Requirements in Regulated Environments

Participants emphasized that **Computerized System Validation (*CSV*)** is essential for any AI tool integrated into regulated preclinical workflows. These systems may be required to demonstrate the ability to pull data consistently and accurately, maintain comprehensive audit trails, and comply with standards such as the OECD principles **[27]** that apply equally to agentic AI tools intended for use in GLP-compliant reporting workflows. While agentic AI tools may not need to be classified as approved diagnostics, they should still undergo structured validation appropriate to their deployment context to enable compliance, reliability, and user trust. In particular, Section 11.10(a) of FDA 21 CFR Part 11 specifically mandates "validation of systems to ensure accuracy, reliability, consistent intended performance, and the ability to discern invalid or altered records."

## 5.3 Integration and Interoperability Challenges

Toxicologic study data is highly fragmented and typically dispersed across multiple platforms, many of which are unique to each institution. These may include LIMS systems, digital image viewers, archived study reports, as well as spreadsheets, PDFs, and custom-built toxicokinetic tools. In addition, organizations often maintain their own proprietary control datasets while also drawing on public repositories. For agentic AI tools to be usable in such environments, they should be designed with flexibility and interoperability in mind. This means either integrating securely with existing platforms through APIs (see **Glossary** for a short description) and connectors, or being deployed behind institutional firewalls, fully aligned with data governance, security, and compliance protocols. Ensuring seamless and secure access to these diverse data sources is foundational for AI tools to deliver meaningful value in toxicologic workflows and requires data capture systems to be interoperable.

## 5.4 Cultural Resistance and Change Management

While attendees expressed openness to innovation, several emphasized that cultural adoption remains a significant hurdle, particularly within large organizations. There are concerns around delegation of control to agents that automate workflows due to the complexity of nuanced study data that pathologists interpret and importance of their role in study reporting. There is also a lingering concern that AI may diminish the role of pathologists, rather than support and amplify their contributions. In many cases, entrenched workflows and institutional inertia can make it difficult to introduce new tools, even when they offer clear efficiency or quality benefits. Additionally, persistent misunderstandings about the purpose of AI (viewing it as automation instead of augmentation) may need to be proactively addressed. Successfully embedding agentic AI into routine practice will require clear communication, strong change management, clear determination of the return-on-investment and visible alignment with the values and expertise of pathologists themselves.

## 5.5 Data Security and Intellectual Property

Given the highly sensitive nature of preclinical study data, participants strongly emphasized the need for stringent data security and deployment controls. There was consensus that agentic AI systems may need to operate within firewalled, project-specific environments, ensuring that data access is restricted and auditable. Attendees stressed the importance of assurances that data would not be leaked, shared externally, or inadvertently used to train third-party or external AI systems. As a result, custom deployment options, whether on-premises or within a virtual private cloud (VPC), were seen as a potential preference for adoption in regulated or proprietary research settings. However, such custom deployment options can limit the agentic AI system's exposure to external data sources, constraining its ability to learn continuously from broader datasets or user behaviors across institutions. This trade-off necessitates a careful balance between local fine-tuning and centralized improvements. To mitigate this, vendors may adopt *federated learning* or *model distillation* approaches, where aggregate learnings (not raw data) from multiple deployments can be shared back subject to appropriate governance, user consent, and data protection agreements. Alternatively, updates to the core model can be made centrally based on synthetic or shared benchmark datasets, with local customization handled via modular plug-ins, institution-specific vocabularies, or context-aware agents. This hybrid model allows continued improvement without compromising data privacy or regulatory boundaries.

Trust in the AI system's data handling protocols is foundational to its acceptance and use. Robust cybersecurity and regular penetration testing may be required to ensure that agentic AI systems, particularly those handling preclinical study data, are protected against unauthorized access, data breaches, and model manipulation.

# 6. Building Trust

Adoption of agentic AI in toxicologic pathology will ultimately depend on **earning and sustaining the trust of its users** – not only through performance, but through clarity, control, and compliance. Participants outlined several foundational principles that may guide the development and deployment of these systems.

## 6.1 Transparency and Explainability

Participants agreed that trust in agentic AI systems begins with explainability. For pathologists and study scientists to rely on AI-generated insights, they should be able to trace how conclusions were derived, including which datasets or literature sources were used and what decision logic guided the interpretation. Outputs should come with citations, confidence scores, and direct links to the underlying data. Importantly, systems should be capable of answering not just *what* was found or *how much* it measured, but also *why* a particular conclusion was reached. Building this level of transparency into AI tools is essential to foster confidence, enable expert oversight, and regulatory defensibility. And finally, there should be guardrails that help the pathologist to consciously criticize the output of agentic AI systems.

## 6.2 Human-in-the-Loop Control

Participants were unanimous in asserting that agentic AI systems should serve as an assistant, not an authority. In toxicologic pathology, it is essential that pathologists retain full editorial control over report content and maintain oversight of any AI-suggested conclusions. The ability to intervene, correct, or reject system outputs at every stage of the workflow was viewed as critical. AI-generated drafts should be treated as efficient starting points, tools that reduce clerical burden and accelerate routine tasks, rather than as final deliverables. This balance between assistance and human oversight is key to safeguarding scientific integrity and maintaining user trust.

## 6.3 Agent Governance and Feedback Loops

For the agentic AI systems to be integrated into toxicologic pathology workflows, establishing robust governance frameworks will be essential. These frameworks include comprehensive audit trails that log agent decisions and outputs, enabling traceability and accountability. As study modalities, regulatory standards, and data ecosystems evolve, these systems will also require ongoing versioning, performance monitoring, and refinement to remain reliable, compliant, and aligned with scientific expectations while adhering to the ALOCA++ principles **[26]**. In addition, in alignment with the OECD AI Principles (2024) **[28]**, the governance of agentic AI systems may require that transparency, traceability, and human oversight are embedded across the system lifecycle, enabling stakeholders to understand, monitor, and meaningfully intervene in autonomous decision-making processes. Equally important are user feedback mechanisms, which allow experts to correct AI outputs and guide ongoing refinement. To enable safety and oversight, there should be clear boundaries on autonomy, specifying which decisions an agent can make independently and which require explicit human confirmation.

# 7. Adoption Roadmap

Implementing agentic AI in toxicologic pathology requires a phased, collaborative, and context-aware approach. The roundtable discussion revealed that success will depend not only on technological capability, but on well-planned engagement with pathologists, data teams, IT teams, histo-technicians, QA, and regulatory leads.

This section outlines a practical roadmap for adoption, as shown in **Figure 2**, starting with focused pilots and building toward broader organizational integration.

## 7.1 Phase 1: Workflow Mapping and Stakeholder Engagement

Before deploying agentic AI tools, a thorough assessment of the existing reporting workflows of user organization may need to be undertaken. This includes conducting process audits to map the entire study reporting lifecycle, from initial slide review to final sponsor sign-off. Within this map, it is important to identify critical decision points, recurring bottlenecks, and data hand-offs where inefficiencies or inconsistencies may occur. Equally essential is engaging directly with key users (such as study directors, study pathologists, toxicologists, research associates, histo-technicians and QA personnel) to understand their current challenges and unmet needs. This groundwork is critical for identifying high-value automation targets and customizing agents' behaviors accordingly. It will ensure that AI deployment is not only technically feasible but also aligned with real-world practices and priorities.

## Adoption Roadmap: From Pilot to Practice

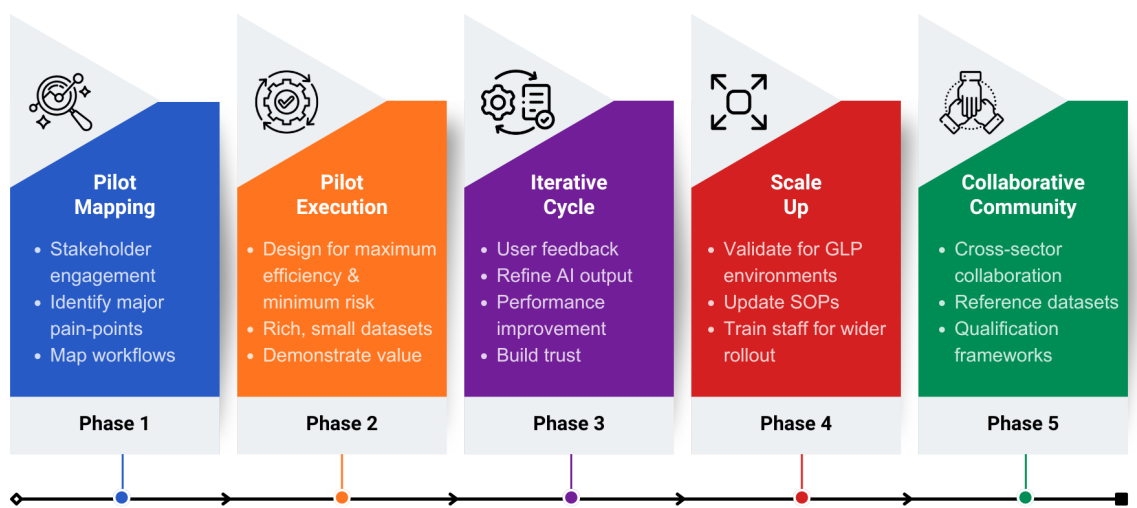

*Figure 2: An adoption roadmap from pilot to wider rollout and community standards.*

### 7.2 Phase 2: Targeted Pilot Studies

In this phase, the goal is to generate meaningful insights while keeping risk to a minimum. Early pilots are best conducted on non-GLP or discovery-phase studies, where regulatory constraints are lower and workflows are more flexible. Ideal pilot datasets include those that are already completed, allowing retrospective evaluation without impacting live projects, and studies with dense data (i.e., with a relatively high volume of lesion diagnoses, complex and subtle findings, multiple dosing regimens, timepoints and endpoints) such as neurotoxicity or lead optimization studies, where AI can demonstrate real value. These pilots should aim to quantify time savings in drafting and quality control, evaluate the interpretability and trustworthiness of AI outputs, and surface any integration or data governance issues that might arise during deployment. This evidence base can help guide broader adoption and refinement.

### 7.3 Phase 3: Feedback, Refinement, and Benchmarking

In this phase, the focus shifts to collaborative iteration between developers and users. After the pilot phase, teams should work together to gather detailed feedback on the system's usability, clarity of outputs, and interpretive accuracy. Performance should be benchmarked against traditional workflows to assess efficiency gains and consistency improvements. Based on these insights, agent behavior, terminology, and report formatting can be refined to better align with institutional standards. Some organizations may also choose to co-develop training datasets or fine-tune the AI models using their own report archives, ensuring the system reflects internal language norms and domain-specific expectations, and enhancing adoption across teams.

To enable downstream validation success, Quality Assurance (QA) teams should be actively engaged by the end of Phase 3, or at the latest early in Phase 4, not merely in an oversight role, but as key contributors to process refinement and alignment, before workflows become fixed in ways that may be difficult to adapt or validate.

### 7.4: Phase 4: Broader Rollout in Regulated Contexts

This phase may be designed to transition an agentic AI system from pilot to production use in GLP environments. Following successful validation, the system may be deployed on a per-study or per-sponsor basis, with oversight from internal QA teams. Continuous monitoring through audit trails and structured user feedback helps maintain performance and regulatory integrity. At this stage, organizations should also implement standard operating procedure (SOP) updates to reflect new human-in-the-loop workflows, provide staff training on how to interpret and edit AI-generated content, and maintain clear documentation to demonstrate compliance with relevant regulatory expectations (e.g., FDA, EMA, and OECD) **[26-28]**. This phase safeguards that the AI system is not only effective but also auditable and regulatory-ready.

### 7.5 Phase 5: Long-Term Vision for Industry Collaboration & Standardization

In envisioning the long-term trajectory of agentic AI in toxicologic pathology, participants underscored the importance of collaborative infrastructure to support trust, adoption, and scalability. Accelerated progress will depend on cross-sector engagement between pharmaceutical companies, CROs, and academic institutions. Key initiatives proposed include the development of: (*a*) shared reference datasets to create public reference datasets to enable benchmarking and foster transparency, and (*b*) qualification frameworks that provide consistent standards for system testing and regulatory alignment. Additionally, participants recommended the formation of or involvement with special interest groups within professional societies and relevant regulatory working groups, to shape guidance, align expectations, and drive collective advancement across the industry.

# 8. Pilot Design Recommendations

Pilots are essential for translating the promise of agentic AI into real-world value. Based on the roundtable discussion, this section outlines practical recommendations for designing impactful and trustworthy pilots, especially in early-phase toxicologic pathology reporting.

## 8.1 Goals of a Pilot

A well-designed pilot can serve as a critical stepping stone toward broader adoption of agentic AI in toxicologic pathology. Its primary aims are to demonstrate measurable time and effort savings in generating study reports, while also rigorously testing the interpretability and transparency of the AI's outputs. Equally important is assessing how well the system fits within existing institutional workflows, minimizing disruption to established practices. Finally, valuable user feedback and insights can guide iterative refinement of the system and inform future rollout strategies.

## 8.2 Potential Use Cases for Pilot Projects

Pilot studies for agentic AI should be strategically designed to balance insight with minimal regulatory risk. Non-GLP environments offer a suitable setting for initial deployment, as they allow for greater flexibility in experimentation and iteration. These contexts are typically less constrained by regulatory oversight, making them ideal for exploring new workflows, adjusting agent behavior, and gathering early user feedback without compliance-related risks. Promising early applications include: study protocol ingestion with automatic metadata parsing, data visualization, automatic selection of relevant findings, correlation of findings, proposals for mechanisms of action, QA review support through intelligent cross-checking of datasets, and automated draft narrative generation for completed studies. These use cases allow organizations to measure efficiency gains in reporting workflows and evaluate value, usability, and integration potential before advancing toward regulated deployment. In GLP settings, appropriate candidates for pilot deployment may include repetitive, high-volume studies such as 13-week rat toxicology or neurotoxicity panels.

## 8.3 Success Metrics

To enable meaningful evaluation, clear and measurable indicators should be defined before launching any pilot of agentic AI in toxicologic pathology. These metrics may include, but are not limited to: the proportion of AI-generated report text accepted without editing, overall time savings in drafting reports compared to traditional methods, and user-reported clarity of the AI's decision logic (for instance, via post-pilot surveys). Additional indicators may include the number of inconsistencies or data gaps flagged by the system, and expert reviewers' trust or confidence scores in the outputs. Finally, success metrics may evolve over time, with the goal of providing a comprehensive view of the system's effectiveness, usability, and potential for real-world adoption.

## 8.4 Roles and Responsibilities

A successful agentic AI pilot in toxicologic pathology hinges on strong cross-functional collaboration, with each stakeholder playing a critical role.

- The study director contributes protocol context and enables consistency in the narrative by reviewing the outputs considering study objectives and standards;
- The pathologist remains central to the diagnosis, responsible for interpretation of histopathology findings, reviewing any AI-generated outputs, and providing expert feedback;
- The lab pre-analytics team, especially to enable delivery of good quality multiplexed or spatial transcriptomics imaging data and to format and double-check the data;
- The AI development team is responsible for configuring the agent workflows, incorporating user feedback, and iteratively improving system performance in response to pilot findings;
- The data engineer or IT lead plays a vital role in enabling access to the necessary datasets and supporting integration between AI tools and institutional systems;
- The QA representative maintains oversight of traceability and ensuring that validation requirements are met, especially in regulated environments.

Together, these roles form the foundation for a robust and trustworthy pilot implementation.

## 8.5 Technical and Ethical Safeguards

To maintain trust and safeguard compliance, pilot studies involving agentic AI may need to be designed with robust safeguards. This includes deploying the system in firewalled or sandboxed environments to prevent unintended data exposure and ensuring that no proprietary data is shared beyond the boundaries of the pilot site.

All AI-generated outputs should be clearly labeled as drafts rather than final interpretations, reinforcing the human-in-the-loop model. Additionally, developers should maintain documented version control and change logs, enabling transparency, traceability, and accountability throughout the pilot lifecycle.

## 8.6 Communication and Documentation

In order to maximize the value and impact of agentic AI pilot studies, teams should adopt a proactive and structured approach to communication and documentation. This starts with sharing clear pilot objectives with all team members in advance, ensuring alignment on goals and expectations. Throughout the pilot, it is important to maintain detailed logs of key decisions, edits to AI outputs, and user comments to support traceability and analysis. Once the pilot concludes, a structured debrief session should be held to gather insights from all stakeholders. Finally, teams should document findings and lessons learned in a concise internal report or presentation, helping to inform broader adoption strategies and future refinements.

# 9. The Path Ahead: Collaboration, Community, and Co-Development

The roundtable discussion made clear that no single institution, vendor, or stakeholder can shape the future of agentic AI in toxicologic pathology alone. Success will require a coordinated effort to **share learnings, establish trust frameworks, align on standards and create interoperable tools**.

This concluding section outlines strategic actions to sustain momentum and shape a shared roadmap.

### 9.1 Establishing a Community of Practice

To foster knowledge exchange and accelerate the responsible adoption of agentic AI in toxicologic pathology, participants emphasized the need for a collaborative, cross-sector forum. Such a platform could enable the sharing of anonymized pilot results, helping stakeholders learn from each other's experiences. It would also help in defining common terminology and clarifying use case categories, fostering alignment across institutions. Importantly, it could serve as a launchpad for joint validation initiatives, reducing duplication and accelerating regulatory readiness. A tangible next step may be to work with existing working groups (such as the IQ Drusafe working group), form or join a working group under a professional society (e.g., STP, ESTP, JSTP) or via a neutral academic partner to provide structure, continuity, and impartial oversight.

### 9.2 Developing Precompetitive Benchmarks

A key enabler for building trust in agentic AI systems within toxicologic pathology may be the development of shared validation infrastructure. This may include the creation of reference datasets, such as publicly available SEND files and curated de-identified study reports that can serve as common benchmarks for testing, although it is recognized that such data is generally not publicly available. Equally important is the establishment of standardized evaluation metrics, jointly defined by CROs, pharmaceutical companies, and regulatory bodies, to enable alignment on what constitutes performance and reliability. Additionally, organizing challenge contests, such as the annual Critical Assessment of protein Structure Prediction (CASP) open contests **[29]** which led to the development of the breakthrough technology AlphaFold **[30]**, that industry and research groups can participate in would provide a structured mechanism for accelerating progress, benchmarking solutions, and community-wide learning. Together, these initiatives would foster transparency, reproducibility, and efficiency, while reducing the burden of duplicative validation across organizations.

### 9.3 Aligning with Regulatory Conversations

As regulatory agencies like the FDA, EMA, and PMDA begin to engage more deeply with the implications of AI in preclinical and clinical workflows, industry stakeholders have a timely opportunity to proactively shape emerging expectations. This can be done by submitting concept papers that outline the role and scope of agentic AI systems in toxicologic pathology reporting, helping regulators understand both their capabilities and limitations. Participation in regulatory sandboxes and workshops focused on AI can foster dialogue, clarify compliance pathways, and identify practical considerations. Furthermore, co-authoring position papers on validation, transparency, and auditability standards can guide policy development and build shared understanding, laying the groundwork for future AI submissions in regulated environments.

### 9.4 Co-Development as a Model

Participants strongly agreed that the most impactful agentic AI tools in toxicologic pathology will emerge from co-development models that actively involve end users. Rather than building solutions in isolation, data scientists and AI vendors should engage pathologists, toxicologists, histo-technicians, IT teams, regulatory and QA teams early in the design process and incorporate iterative feedback loops throughout the development lifecycle in order to enable alignment with real-world workflows, expectations, and terminology. Flexibility is also crucial; systems should allow for institution-specific fine-tuning or modular adaptation, so they can integrate seamlessly into local practices. Finally, successful collaboration depends on clear IP, commercial and data use agreements that protect institutional data ownership while recognizing and rewarding the contributions of AI experts.

### 9.5 From Reporting Tool to Scientific Assistant

Looking to the future, participants envisioned agentic AI systems evolving to serve as generalist research assistants across the toxicologic pathology and drug development landscape. Such systems could help generate mechanistic hypotheses, support exploration of adverse outcome pathways (AOPs), and synthesize literature to inform study design and compound selection. As well as accelerating study reporting, the broader opportunity is to enable richer scientific insight, enhanced decision-making, and more robust experimental design achieved through a collaborative partnership between human experts and agentic AI systems.

# Acknowledgements

The authors would like to thank Dr Aleksandra Zuraw (Charles River Laboratories) for their thoughtful feedback and insightful comments on an earlier draft of this manuscript.

# Disclaimer

*The views and opinions expressed in this white paper are solely those of the authors and do not necessarily reflect the official policy or position of their respective employers or affiliated organizations.*

# Appendices

## A1. Glossary of Terms and Concepts

| Term | Definition |
|---|---|
| **AI (Artificial Intelligence)** | Computer systems that perform tasks typically requiring human intelligence, like pattern recognition or decision-making. |
| **Traditional AI** | Rule-based or narrow AI designed for specific tasks using structured data and fixed input formats. |
| **LLM (Large Language Model)** | An AI model trained on large text datasets to understand and generate human-like language. |
| **AI Agent** | A software module powered by AI (often LLM-based) that performs a specific task within a workflow. |
| **Agentic AI** | AI systems capable of multi-step reasoning, retrieving and synthesizing data across sources, and performing tasks with human oversight. |
| **API (Application Programming Interface)** | A standardized software interface that allows different systems or applications to communicate and exchange data securely and efficiently. Commonly used to connect AI tools with platforms like LIMS or image viewers. |
| **Data Ingestion** | The process of collecting, importing, and harmonizing structured or unstructured data from multiple sources (e.g., LIMS, SEND, image archives) into a format usable by an AI system. |
| **Contextualization** | The process of linking findings or data points to their broader scientific, study-specific, or regulatory context to support interpretation and decision-making. |
| **Evidence Integration** | The synthesis of insights from diverse datasets to support contextualized interpretation and decision-making. |
| **Narrative Drafting** | The process of generating written summaries of study results, often supported by AI, in tox path workflows. |
| **Human-in-the-Loop** | A system design where humans retain oversight and can correct or override AI outputs. |
| **Out-of-the-Box Performance** | How well an AI tool performs using default settings, without customization. |
| **On-Premise Deployment** | Hosting AI systems within the physical infrastructure of the user organization for full data control. |
| **VPC (Virtual Private Cloud)** | A secure, cloud-based but isolated environment under user control for safe AI tool deployment. |
| **Confidence Scores** | Indicators (numerical or qualitative) reflecting an AI system's certainty in its outputs. |
| **Interpretability** | The degree to which humans can understand how an AI system reached its output or conclusion. |
| **QA (Quality Assurance)** | Ensures regulatory compliance and data integrity in GLP workflows, including validation and audit processes. |
| **Tuning Dataset** | A data subset used to adjust model parameters during development—distinct from validation datasets in regulated contexts. |
| **Testing Dataset** | Used to assess an AI system's performance post-development; not used in training or tuning. |
| **Validation** | Demonstrating that a computerized system consistently performs as intended in its specific deployment. |
| **Audit Trail** | A secure, time-stamped log of system actions and decisions. |
| **CFR Part 11** | FDA regulations governing electronic records and signatures in regulated environments. |